\documentclass[british,12pt]{article}

\usepackage{babel}

\usepackage{graphicx} 
\graphicspath{{figures/}}

\usepackage{xcolor}
\pagecolor{white} 

\title{\textbf{\textsf{JENA White Paper on \\   \vspace{0.3cm} 
  \Huge{Software and Heterogeneous Architectures}}}}
\date{} 

\usepackage{amsmath, amssymb}

\usepackage[style=numeric-comp,sorting=none,backend=biber,language=british]{biblatex}
\usepackage{helvet}
\usepackage{sectsty}
\allsectionsfont{\sffamily} 

\usepackage{geometry}
\usepackage{hyperref} 
\usepackage{color}    
\definecolor{pigment}{rgb}{0.2, 0.2, 0.6}
\hypersetup{
  colorlinks = true, 
  urlcolor   = pigment, 
  linkcolor  = black, 
  citecolor  = pigment 
}

\usepackage{fancyhdr}
\usepackage{tcolorbox}
\newtcolorbox{examplebox}{
  colback=white,
  colframe=gray!30,
  title=Recommendation,
  sharp corners,
  boxrule=0.5pt,
  coltitle=black
}

\usepackage{ifthen}
\newboolean{showinstructions}
\newboolean{showexamples}
\newboolean{showexplanations}

\newcommand{\explanation}[1]{%
  \ifthenelse{\boolean{showexplanations}}%
    {\textit{Explanation:} #1}%
    {\ignorespaces}%
}

\newcommand{\instructions}[1]{%
  \ifthenelse{\boolean{showinstructions}}%
    {#1}%
    {\ignorespaces}%
}

\makeatletter
\newcommand{\maketitlepage}{%
    \begin{titlepage}
        \maketitle
        \thispagestyle{empty}
        \vfill 
        \centering
        \includegraphics[width=0.9\textwidth]{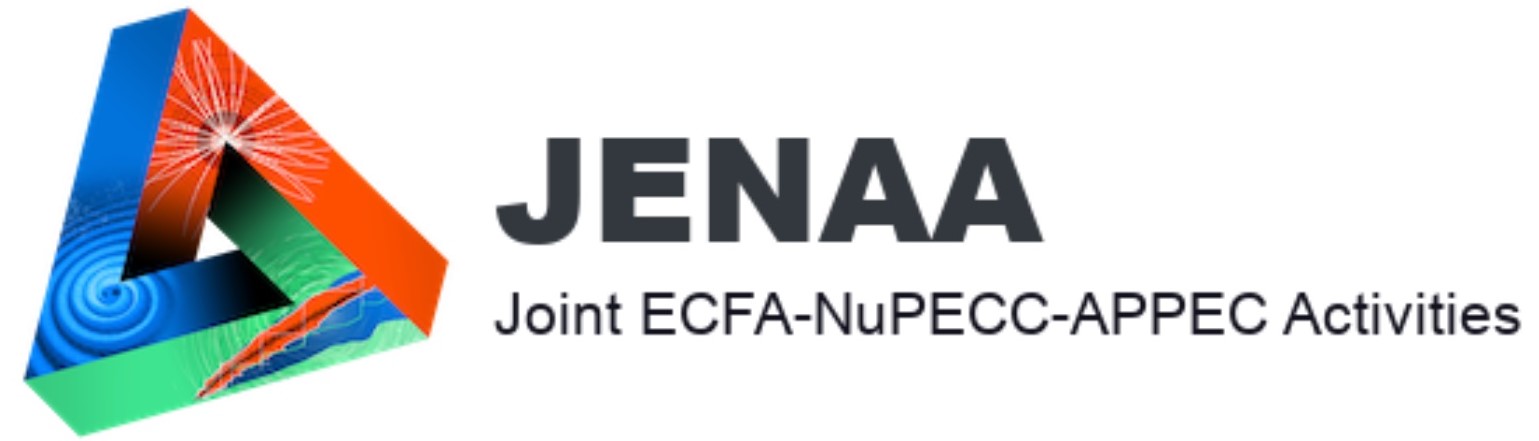} \\ 
        \vspace{0.7cm}
        \includegraphics[width=0.9\textwidth]{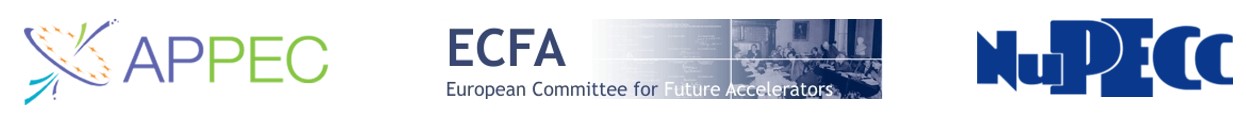}
        \vfill
    \end{titlepage}
    \newpage
}
\makeatother

\usepackage{authblk}
\usepackage{orcidlink}

\author[1]{\small Mohammad Al-Turany~\orcidlink{0000-0002-8071-4497}}
\author[2]{David Chamont~\orcidlink{0000-0003-2618-7355}}
\author[3]{Davide Costanzo~\orcidlink{0000-0003-4920-6264}}
\author[4]{Caterina Doglioni~\orcidlink{0000-0002-1509-0390}}
\author[5]{Håvard Helstrup~\orcidlink{0000-0002-9335-9076}}
\author[6]{Bruno Khélifi~\orcidlink{0000-0001-6876-5577}}
\author[7]{Thomas Kuhr~\orcidlink{0000-0001-6251-8049}}
\author[8,9]{Paul Laycock~\orcidlink{0000-0002-8572-5339}}
\author[10]{Adrien Matta~\orcidlink{0000-0002-0331-1563}}
\author[11]{Eva Santos~\orcidlink{0000-0002-0474-8863}}
\author[12]{Luis Sarmiento Pico~\orcidlink{0000-0001-7600-2772}}
\author[13]{Fabien Schüssler~\orcidlink{0000-0003-1500-6571}}
\author[12]{Oxana Smirnova~\orcidlink{0000-0003-2517-531X}}
\author[14]{Graeme A Stewart~\orcidlink{0000-0003-0182-7088}}
\author[15]{Gabriel Stoicea~\orcidlink{0000-0002-7511-4614}}
\author[16]{Liliana Teodorescu~\orcidlink{0000-0002-6974-6201}}
\author[17]{Christoph Weniger~\orcidlink{0000-0001-7579-8684}}

\affil[1]{GSI Helmholtzzentrum für Schwerionenforschung, Germany}
\affil[2]{IJCLab, France}
\affil[3]{University of Sheffield, UK}
\affil[4]{University of Manchester, UK}
\affil[5]{Western Norway University of Applied Sciences, Bergen, Norway}
\affil[6]{CNRS/IN2P3, France}
\affil[7]{Ludwig-Maximilians-Universität München, Germany}
\affil[8]{Département d'Astronomie, Université de Genève, Chemin Pegasi 51, Versoix, Switzerland}
\affil[9]{Gravitational Wave Science Center (GWSC), Université de Genève, Geneva, Switzerland}
\affil[10]{Université de Caen Normandie, ENSICAEN, CNRS/IN2P3, LPC Caen UMR6534, F-14000 Caen, France}
\affil[11]{FZU - Institute of Physics of the Czech Academy of Sciences, Czech Republic}
\affil[12]{Lund University, Sweden}
\affil[13]{IRFU, CEA, Université Paris-Saclay, F-91191 Gif-sur-Yvette, France}
\affil[14]{European Organisation for Nuclear Research, CERN, Switzerland}
\affil[15]{IFIN-HH, Romania}
\affil[16]{Brunel University of London, UK}
\affil[17]{Amsterdam University, Netherlands}

\begin{document}

\maketitlepage

\pagenumbering{roman}
\vspace*{0.5cm}

\tableofcontents
\thispagestyle{empty}

\vspace{7cm}



\pagenumbering{arabic}
\setcounter{page}{1}

\section{Executive Summary}

The scientific communities of nuclear, particle, and astroparticle physics are continuing to advance and are facing unprecedented software challenges due to growing data volumes, complex computing needs, and environmental considerations. As new experiments emerge, software and computing needs must be recognised and integrated early in design phases. This document synthesises insights from ECFA, NuPECC and APPEC, representing particle physics, nuclear physics, and astroparticle physics, and presents collaborative strategies for improving software, computing frameworks, infrastructure, and career development within these fields.

\section{Introduction}

Across the scientific domains represented by the APPEC, NuPECC, and ECFA organisations there are very significant challenges in software and computing, driven by ambitious physics programmes that deliver new detectors and observatories with increased data rates and data complexity. We discuss in this white paper these challenges as related to the software that is used directly to produce and process our science data, and to operate the corresponding infrastructure. The software to support these instruments, which is often very specific, is frequently ageing and needs investment, or replacement. This is particularly challenging for smaller experiments, where any dedicated effort is difficult to find within a smaller team. As well as data rates, the need for dedicated software effort is driven by a necessity to adapt to use modern hardware platforms, where high levels of parallelism are needed, particularly to execute efficiently on devices such as GPUs. The observational data, along with simulated data that models physics and detectors, needs to be reconstructed, analysed, increasingly in a distributed context, across sites and utilising facilities such as HPCs. It must be made available according to FAIR principles, which brings additional costs to genuinely achieve open science for these experiments. As software is so essential for our science domains, it is imperative to improve the recognition of those developing software and to support their careers.

In this paper we review the upcoming challenges, identify best practices, and recommend how to bolster support for widespread adoption of effective solutions to the needs of ECFA, NuPPEC and APPEC.

\section{Computational and Data Management Challenges}

\subsection{Growing Data Volumes}

\begin{figure}[htbp]
    \begin{center}
        \includegraphics[width=0.7\textwidth]{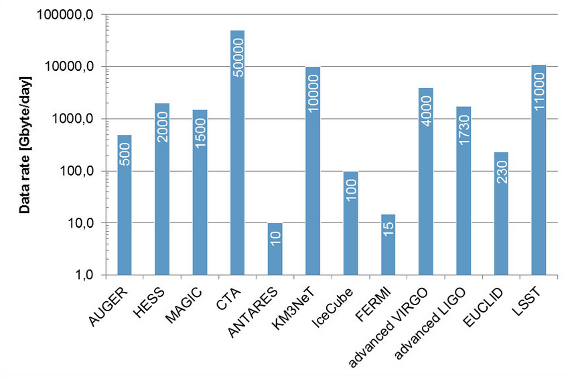} \\ 
        \caption{Data rates (Gbytes/day) for several astrophysics experiments~\cite{berghöfer2015modelcomputingeuropeanastroparticle}.}
        \label{figures:Fig1}
    \end{center}
\end{figure}

\begin{figure}[htbp]
    \begin{center}
        \includegraphics[width=0.49\textwidth]{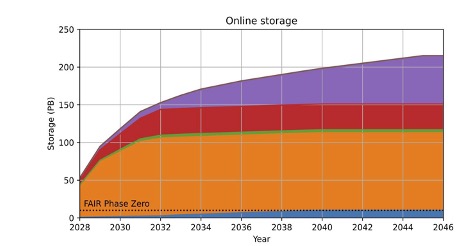} 
        \includegraphics[width=0.47\textwidth]{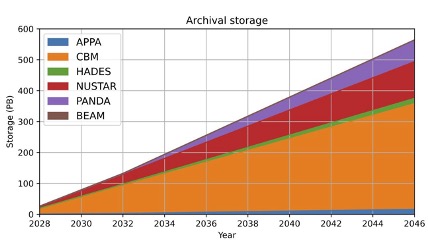} \\ 
        \caption{The required amount of storage as a function of year for the FAIR experiments at GSI. The left panel depicts the requested disk space for fast access, whereas the right panel presents the needed long-term storage space (archive). The contributions of the various research lines are indicated by different colours. The dashed line shows the current storage for FAIR Phase Zero activities.} 
        \label{figures:Fig2}
    \end{center}
\end{figure}

\begin{figure}[htbp]
    \begin{center}
        \includegraphics[width=0.45\textwidth]{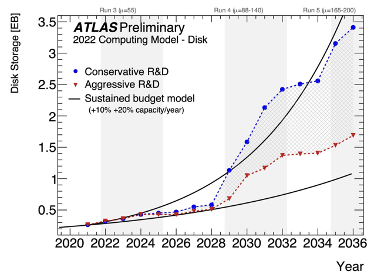} 
        \includegraphics[width=0.45\textwidth]{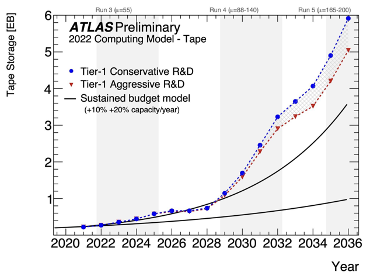} \\ 
        \caption{ATLAS projected evolution in exabytes of disk (left) and tape (right) usage from 2020 until 2036, under the conservative (blue) and aggressive (red) R\&D scenarios~\cite{CERN-LHCC-2022-005}. The grey hatched shading between the red and blue lines illustrates the range of resources consumption if the aggressive scenario is only partially achieved. The black lines indicate the impact of sustained year-on-year budget increases, and improvements in new hardware, that together amount to a capacity increase of 10\% (lower line) and 20\% (upper line). The vertical shaded bands indicate periods during which ATLAS will be taking data (not yet updated to the latest HL-LHC planning). Similar plots for the CMS experiment are available~\cite{Software:2815292}.}
        \label{figures:Fig3}
    \end{center}
\end{figure}

Figures \ref{figures:Fig1}--\ref{figures:Fig3} show the data challenge for the JENA communities, which are already reaching the exabyte scale, with still significant growth to come. While some communities, notably the LHC experiments, have dealt with such extreme data volumes for some time, petabyte-size datasets are now a de facto specification of modern scientific data instruments.  These large data volumes represent a significant challenge to the scientific communities that need to manage and analyse them.

\subsection{FAIR Data Management}

Beyond the basic provision of computing hardware to store the data, one of the biggest difficulties for scientific communities is to manage their data according to the FAIR principles~\cite{Wilkinson2016}. In particular for smaller experiments, where the short lifetime of an experiment combines with the diversity and uniqueness of the metadata involved, this presents significant challenges.  While larger experiments, starting with the LHC experiments, have converged on Rucio~\cite{Barisits2019} as a data management solution, the complexity of this solution still represents a significant obstacle to smaller teams of scientists who have insufficient time and computing expertise to leverage it (there are other efforts targeting smaller communities, such as EUDAT~\cite{EUDAT}, but these have not yet seen adoption in ENA). Even though Rucio supports FAIR data principles, it requires significant expert effort to provide sufficient metadata to Rucio to fully adhere to those principles, with additional work needed to make data available outside of a collaboration.

\subsection{Growing Compute needs}

\begin{figure}[htbp]
    \begin{center}
        \includegraphics[width=0.8\textwidth]{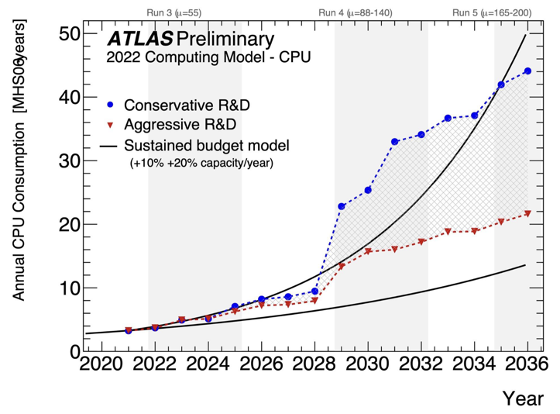} \\ 
        \caption{ATLAS projected evolution of compute usage from 2020 until 2036, under the conservative (blue) and aggressive (red) R\&D scenarios~\cite{CERN-LHCC-2022-005}. The grey hatched shading between the red and blue lines illustrates the range of resources consumption if the aggressive scenario is only partially achieved. The black lines indicate the impact of sustained year-on-year budget increases, and improvements in new hardware, that together amount to a capacity increase of 10\% (lower line) and 20\% (upper line). The vertical shaded bands indicate periods during which ATLAS will be taking data  (not yet updated to the latest HL-LHC planning). Similar plots for the CMS experiment are available~\cite{Software:2815292}. }
        \label{figures:Fig4}
    \end{center}
\end{figure}

The growing data volumes also often contain more information per unit volume, and this increasing complexity translates into an increased need for computing power.  Figure \ref{figures:Fig4} shows the example of the ATLAS experiment~\cite{CERN-LHCC-2022-005} (for CMS see~\cite{Software:2815292}), where even with aggressive R\&D the CPU resource consumption may grow beyond the increased capacity provided by year-on-year budget increases and hardware improvements.

Overall, when surveyed, 90\% of respondents to the JENA-Spectrum survey reported that improving their software performance is an issue~\cite{JENA-Spectrum-Survey-Report}.

\subsection{Distributed computing}

The increased need for computing power benefits from distributed computing solutions that allow scale-out beyond the capacity of a single computing facility. Again, the LHC experiments have developed and optimised distributed computing solutions for some time (including significant integration into commercial cloud resources~\cite{Megino_2024}).

For large experiments with reasonable numbers of computing experts, the benefit of using these solutions can be transformative. Unlike Rucio, there is less convergence on a common solution even within the LHC experiments. This itself represents a problem, as the communities supporting these solutions are thereby smaller, and small experimental teams have difficulties in understanding the difference between these complex solutions. In addition, as with Rucio, adoption of a solution requires a significant amount of ongoing effort from people with sufficient computing expertise.

\section{Technology and Hardware Evolution and Challenges}

\begin{figure}[htbp]
    \begin{center}
        \includegraphics[width=0.7\textwidth]{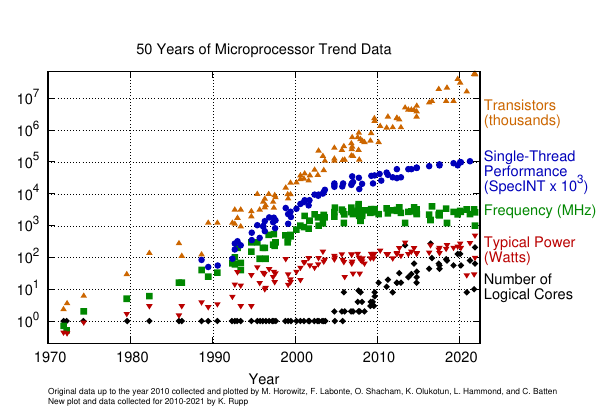} \\ 
        \caption{Microprocessor trends over the last half century, from~\cite{KRupp_Microprocessor_Data}.}
        \label{figures:Fig5}
    \end{center}
\end{figure}

The hardware technology used to process data has undergone significant changes in the last decades. CPU clock frequencies largely stopped increasing since the mid-2000s (see Figure \ref{figures:Fig5},~\cite{KRupp_Microprocessor_Data}). This has drastically limited improvements in single core performance. Moore's Law has, however, continued, so that greater and greater numbers of transistors are found on CPU dies. These transistors are used to increase the number of cores available on the chip and to add wide vector registers that support single instruction multiple data processing (SIMD).

A related innovation is the development of the general purpose processing capabilities of graphical processing units (GPUs). These are massively parallel processors, often with thousands of cores, which can achieve extremely high performance on classes of problems where the same operation is performed on wide banks of data. Data access patterns and predictability of code flow are critical to achieve good performance and not all problems are easily adapted to this model. Data must be transferred from the CPU host to the GPU device, so latency is critical to control, usually achieved by ensuring that large amounts of data are transferred and that the GPU constantly has work to do. It should be noted that the evolution of GPUs is very driven by machine learning at the moment and, while this is something we also can take advantage of (see JENA WP4 \cite{JENA_WG_Reports}), it is favouring the development of processing power for low precision floating point operations. Thus, double precision calculations, which are often what we need, are hardly improving at all.

Architectural divergence is another significant trend. After many years where x86\_64 machines dominated, ARM processors are now making significant inroads into data centres~\cite{ARM_Datacentre}. In the JENA survey 27\% of survey respondents have software supported on AArch64. Here they compete favourably in power consumption per unit of work. Other architectures on the horizon include RISC-V, where Europe has strategic ambitions. All of this requires software which supports these platforms. For GPUs the situation is also diverse, with Nvidia dominant~\cite{Datacentre_report}, but with hardware from AMD and Intel available -- JENA communities already target these platforms: 48\%, 21\% and 17\% of our survey reported building software for these platforms, respectively. Note that all of these devices have SDKs which are different, so generic coding for GPUs is still problematic. 

For real-time data processing FPGAs also play a role, as these devices excel at high throughput and low latency applications. The challenge here is the steep learning curve for firmware programming, a task that is generally executed by specialised engineers. Packages that convert code from other languages to FPGA programming languages (e.g., VHDL) have been developed within and outside the field, a widely used example for ML applications is HLS4ML~\cite{Duarte:2018ite}.

Finally, a totally different computing paradigm emerges in quantum computing. This is still in its infancy, and delivery of usable quantum computers is on an unknown timescale (and probably at least a decade away). When (or if) they arrive, this will mean utterly different programming models for which much R\&D would be required. It should also be noted that quantum computing would supplement current digital computers, not replace them.

\section{Software Development and Best Practices}

High data rates, data complexity, challenging hardware, and the need to adapt to the latest algorithmic techniques all demand improved software development practices. Here we summarise some of the most important trends that have emerged. All of these require effort and appropriate skills. Adoption remains patchy across our communities, pointing to a lack of investment in these areas. Recently funded initiatives, such as EVERSE~\cite{EVERSE}, have begun their work on delivering frameworks and recommendations that researchers in our fields can adopt to improve their software development, as in ``Guidelines'' below. The FAIR4RS~\cite{Barker2022} principles are supported by almost all of the following points.

\subsection{Open Source Software}

Initiatives on open software started in the 80s, and the three JENA communities made a significant effort to adopt open source policies since the 2010s, making sure research software source code was made public. However, making software open source does not mean it can be used by others and, shockingly, 46\% of survey respondents (mostly individual developers and small groups) reported not having a licensing model for their software.
Clear software policy on copyright and licences is essential to allow effective scientific collaboration and the community is already widely using licences such as Apache 2 or GPL. There should be strong advice to stick to standard licences and to understand the implication of a particular licence choice, including incompatible choices. We note that HSF~\cite{jouvin_2016_1469636}, CERN~\cite{Fluckiger:1482206} and GSI/FAIR~\cite{GSIFAIROpenSource} have published guidelines. 

\subsection{Software versioning}

Software versioning is a cornerstone of code management, and widespread adoption by the community leads to a noticeable improvement in software quality~\cite{swebok4}. The practice is now adopted by most software collaborations, even the smallest ones. This has been possible with the widespread adoption of git as a version control system (replacing less good solutions, such as SVN) and the deployment of institutional code versioning services, such as Gitlab, or Microsoft's GitHub. This is a good example of how institutional investment makes a noticeable difference in scientific output.

Code versioning helps guarantee that the software is preserved. Code versioning sites help to make it publicly available, and promote improved quality by providing tools for issue tracking, code review, documentation, and CI/CD, to mention but a few.

\subsection{Automated workflows}

Automation of complex software workflows, such as data analysis and simulation, is key to mitigate the impact of resource intensive computing tasks~\cite{swebok4}. In addition, an automated workflow provides an excellent means towards reproducible analysis, by encoding the whole production process of a scientific result in a machine, and sometimes human-readable way. Platforms such as REANA~\cite{REANA} aim at delivering a comprehensive environment to deploy such workflows effectively. At the moment use in our communities is patchy and should be expanded.

Software should be developed with the use of workflow automation tools, such as Snakemake~\cite{snakemake} or DVC~\cite{The_DVC_team_and_contributors_DVC_Data_Version}, in mind -- in particular to be properly versioned, as mentioned above. The crucial addition of data versioning and provenance should also be integrated in those workflows. 

\subsection{Deployment and distribution}

Many paths to deployment and distribution exist and suitable choices should be made depending on the use case. The technology is also evolving at a fast pace and exploring new options and then adjusting best practice is essential. Expert networks, such as EVERSE, are an important tool for this technology watch and dissemination activity. 

While source code distribution should always be available, in practice it rarely allows for easy execution of the software. Recording the dependencies and the environment of the software (e.g., using Conda and similar tools) is an essential first step towards a more complete software distribution.

In addition, offering the application as a ready to use container (e.g., Docker, Apptainer) is an excellent choice both for deployment and long term preservation and reproducibility. Code versioning front-ends offer solutions to automate the production and distribution of containers, however this can become resource-intensive for the computing centre providing the service. For instance IN2P3's Gitlab instance no longer supports building containers on its servers and asks collaborations to provide the computing resources to produce their own containers.

Again, maintaining the required investment in person power, training and infrastructure is crucial to continue offering the tools and services needed to apply these good practices.  The level of investment needs to be continually monitored to ensure that it is maintained.

\subsection{Documentation}

Appropriate documentation should be provided with all software, regardless of its size and use. It is important to differentiate the type of documentation that should be provided: user documentation on how to install and run the software; scientific documentation explaining what the code does and how; and, finally, developer documentation, on how the code is structured and the functionality implemented.

An extremely promising approach, which would reduce the burden on experts, is to use AI to help generate docstrings for code; more adventurously, training large language models on external packages, user code (for how it is really used), and posts on user help forums could provide a modern expert AI-based help system. It will require some investment to make  effective use of these tools and integrate them into scientific ecosystems.

In any approach, documentation activities must be better recognised and rewarded, in order to improve developer motivation on this subject.  Undocumented software negatively impacts the ability of collaborations to deliver their science in the medium and long-term.  The implications for careers in scientific computing are discussed in JENA WP5 \cite{JENA_WG_Reports}.

\subsection{Software collaboration and Management}

In many instances software is developed as part of a detector system or experiment. While this approach allows each collaboration to limit its software dependencies, it has the drawback of limiting the scientific impact of the software (due to limited re-use) and limiting the accessible pool of software specialists.

Of course exceptions exist, with software such as ROOT~\cite{Brun1996}, GEANT4~\cite{Agostinelli2003}, CORSIKA~\cite{Heck:1998vt}, etc. ubiquitous within our three communities. Other examples are middleware which provides key functionality to manage and access storage and computing resources, and the use of smaller toolkit components, e.g., AwkwardArray~\cite{Pivarski_Awkward_Array_2018}, that leverages the flexibility of the Python ecosystem in a very domain independent way.

Strengthening software collaborations should help the community to have a clearer picture of what could be reused and how, favouring cross-collaboration development. Moreover, such well structured collaborations should mean they have a longer lifetime, supporting better documentation, optimisation and support. This all requires good software management, supported by Software Management Plans~\cite{SMPs}, with a clear understanding of costs (including the very real cost of poor software), value and recognition.

\subsection{Legacy Code Maintenance and Modernization: toward continuous integration}

Many software suites and tools are based on legacy code, which poses challenges in terms of maintenance and compatibility with new technologies. This raises the question of what strategy to adopt to maintain those crucial tools. Either training new developers to take over the maintenance of those tools (e.g., training in Fortran, which is still used quite widely, particularly in the theory domain) or redeveloping that software in more modern languages, will require some significant investment. Rewrites will usually be more expensive, but offer considerable opportunities to optimise the software, so would be more justified for critical code. We have many examples of critical software ported from Fortran to C++: PAW to ROOT, GEANT3 to GEANT4, Pythia6 to Pythia8~\cite{10.21468/SciPostPhysCodeb.8}. Those investments proved to be essential in maintaining and extending our software capabilities in the long run.

This in turn implies the need for a clear software road map, or software management plan, at the institutional level. This would list critical software for our communities and make sure that resources and skills are matched to guarantee its maintenance, including necessary updates, and optimisation.

\subsection{AI and ML use in software}

The increasing use of AI and ML techniques within our software suites has opened new doors to software development, including AI coding assistants, such as Copilot. These opportunities are still evolving quickly and it is therefore very difficult to define a set of established good practices. However, there is clear potential to improve developer productivity.

There is also the question of integrating AI/ML workflows into the software itself. For example, MLOps, a suite of services allowing the tracking and versioning of models, is essential for reproducibility and optimisation. Those services are not yet mainstream at the institutional level, but this should be anticipated, just like DevOps services are a cornerstone for software development today. To some extent DevOps and MLOps start to interact and overlap, with Gitlab offering Model Registry and MLflow interfaces.

It is clear that these novel techniques could allow new types of code optimisation and therefore have a very positive impact on computing needs in the future. In particular, AI-driven simulation or AI monitoring of code could help save large amounts of computing resources. This subject is covered in much more detail in JENA WP4 \cite{JENA_WG_Reports}.

\subsection{Performance and Portability}

As discussed in the section on hardware evolution, developing high performance scientific software is a real challenge in a heterogeneous environment. Effort to avoid architecturally specific code is not too onerous for CPU applications, but it remains challenging for GPUs or FPGAs, where generic SDKs lag behind vendor specific offerings and evolution is unclear~\cite{atif2023evaluatingportableparallelizationstrategies}. Vendor lock-in is to be avoided and the academic community should push for more open standards, e.g., the Alpaka library~\cite{MathesP3MA2017} and evolution of C++ itself. However, for performance critical code (e.g., real-time data analysis and reduction), where such architectures are needed, significant skilled developer effort is required. Overall, 90\% of survey respondents reported that software performance has to improve in the future.

In all cases, validation systems need to be enhanced to cover multiple architectures that will not be bit-for-bit identical; and code distribution for multiple architectures adds complexity.

\subsection{Guidelines}

Best practices should be made clear at the institutional level. Guidelines are essential as are effective tools that can be used to establish a common ontology, a clear definition of best practices, and how best to implement them. 

A few instances of such guidelines exist~\cite{RSQkit}, although it is still not a widespread practice. Guidelines should stay flexible and be updated regularly by software specialists. They should propose solutions that are scalable and sustainable, i.e., simple solutions for smaller collaborations with more complex ones reserved for larger collaborations, where investment is amortised~\cite{australian_research_data_commons_2022_6378082}.

This is an important step towards recognition of the value and impact of software. The threshold for publication should be made clearer, as well as the means to publish.

\subsection{Software publication}

It should be made clear that software should be considered a research output in itself, and more regularly published as an academic work. Best practice is to have a reference paper in a refereed journal, including modern lightweight software publications such as Journal of Open Source Software~\cite{JOSS}, combined with an automated archival process to Zenodo and/or Software Heritage. Both are compatible, and then give DOIs that cover the software in general and specific versions that can be cited and the code itself becomes available via its DOI. This helps address the Findable and Accessible elements of the FAIR(4RS) principles. The recent ESCAPE OSSR initiative makes use of Zenodo to provide a framework for well defined metadata and quality checks from the community to make sure referenced software is readily reusable and meets FAIRness criteria.

The community of software specialists also needs to contribute as reviewers in both specialist journals (e.g., JOSS) and curation services (e.g., ESCAPE OSSR) to ensure the quality of publications.  This curation work, crucial to making software FAIR, must in turn be recognised and accounted for.

\section{Open Science: the FAIR Principles}

There is a large overlap in the requirements of meeting FAIR data and software principles
and the requirements of Open Science. To support Open Science, scientific experiments need to, in many ways, operate as observatories, producing well curated data products for direct consumption by anybody, and support this data with suitable software. This is directly the role of the EU science clusters and various other EOSC initiatives.

\subsection{Open Science and FAIR Data}

Independently of embargo periods, the requirement to provide the data for Open Science fulfils FAIR data principles.  Although the huge and complex raw data volumes produced by LHC experiments do not lend themselves to Open Science, much progress has been made in producing smaller, curated datasets that do.  Meanwhile, small and medium-sized experiments, thanks to the challenges of applying FAIR principles, arguably have the hardest time in producing data that can be used for Open Science.

The convergence upon Rucio as part of a FAIR Data Management solution is encouraging.  Development is still needed, particularly in the realm of metadata, which can be very challenging.  Without all of the accompanying metadata needed to completely exploit the scientific data products, the scope of Open Science is limited.  Limitations for Open Science correlate strongly with weaknesses in data preservation efforts, and thereby suggest potential concerns over the long-term curation of the data.  Equally, if these problems can be addressed by Rucio, and the various scientific communities are given the resources they need to adopt Rucio, the outlook is positive.

\subsection{Open Science Frameworks and FAIR Software}

Reproducibility is a key requirement for science.  It is therefore surprising, at least for non-experts, to discover that reproducibility (in the strict sense of the word) is in practice often extremely difficult to guarantee, especially for the long term. A large part of the problem stems from the way that much of the scientific software used to produce results is developed and maintained (see section 4).

Frameworks like REANA (a platform for Reusable and Reproducible Analysis) aim to address this problem by encapsulating an entire analysis.  Analysis typically consists of a series of steps that constitute a workflow.  REANA uses a combination of containers, to capture the whole software stack for each step, and a workflow language, e.g., Snakemake, to express the workflow including all necessary metadata. Publishing the software in containers together with the complete workflow including metadata and provenance is a powerful step towards FAIR software.

\subsection{FAIR Principles and Open Science}

The synergies between the requirements of Open Science, and fulfilling the requirements of FAIR Data and Software, should be leveraged.  Supporting Open Science directly implies the application of FAIR principles which is of direct benefit to the experiments themselves. This requires experiments to have the resources they need to exploit the required tools, requiring investment from funding agencies.

\section{Environmental Sustainability and Resource Allocation}

Scientific computing expends environmental resources. This includes equipment manufacturing costs (CO2, rare earth metals, purified water, etc.) as well as the costs of powering machines in data centres. Long term storage costs for data should also not be neglected.

First and foremost in the strategy to reduce environmental costs is to avoid performing data processing and computation. While this might seem a facile point, critically examining the value of a particular processing step for its scientific benefit should be done regularly.  For example, as experiments mature and detectors are better understood, calibration workflows can be optimised allowing very large data formats to be dropped. Secondly, software itself must be efficient at its job -- this should not come at the expense of maintainability, but opportunities (the famous ``3\%''~\cite{10.1145/356635.356640}) should not be neglected. This implies strongly that ``heavy lifting code'', which is numerically intensive, has to live in a language such as C/C++, Fortran or Julia; use of Python should only be at higher levels. Common or reusable software can unlock the development effort required for sustainable efficient codes, particularly for smaller experiments who could not develop this independently.

Data reduction should be tackled as early as possible in the process for two reasons. Firstly, it allows for less data to be saved to long term storage, reducing costs. Secondly, it allows for early extraction of the most useful physics results at higher rates. This can improve physics reach and, in the case of astroparticle physics, is also an essential component of science alerts for multi-wavelength prompt observations.

Ensuring that an experiment or observatory can use the most optimal hardware for its workflow is also important, given the heterogeneous nature of future scientific computing. This would imply that software is built for x86\_64 and ARM processors, and potentially also on GPUs (see 4.10). This is a software and operational challenge that requires investment in good software engineering and validation. The benefit is enhanced flexibility to run where resources are available. For smaller experiments, lacking expert effort, use of generic and toolkit solutions can still help them to benefit.

To really take advantage of this kind of flexibility, it is necessary to invest in a good data and workflow management system, which can access available resources, managing the needed data. For this to be viable for smaller projects, development of tools that are easier to deploy and manage is needed, as well as laboratory support for operations. This is a general problem for scientific computing where investment in suitable tools can help a lot. Good workflow management (e.g., tools like Snakemake) can also optimise which parts of a workflow need to be rerun, avoiding duplicate jobs and reducing computing costs.

While it is somewhat outside the scope of a software discussion, computing sites can also contribute directly to reducing monetary and environmental costs. Power efficient architectures, low PUEs and heat recovery can help minimise impact. Power can be sourced sustainably and power loads reduced when energy supplies become carbon intensive. This requires dialogue between resource providers and users so that expectations and turn-around time are understood.

\section{Human Resources, Training and Career Development}

\subsection{Bridging the Skills Gap}

The community is simultaneously facing a skills crisis, with a lack of human resources, and a growing complexity of software and computing tools, which absolutely demands specialists. Bridging this gap is difficult because the domain is in direct competition with a strong and wealthy industry that offers high quality and high salary jobs. The situation calls for a rethinking of the career paths of software specialists, whether from a software engineering or physics background.

Training and hiring of teams of multi-skilled staff with both domain knowledge and software skills is essential. This requires investment from the community to bring early career staff to a high level of competency. At the same time, institutions should develop and enforce a strong hiring policy offering permanent positions early to promising software specialists.

Two main career paths could be identified: on the one hand, software specialists hired primarily as physicists, we might say \emph{domain specific software experts} (DSSEs), bringing insight on what the community needs; on the other hand, software experts hired in engineering positions, as \emph{Research Software Engineers} (RSE), who bring expertise on the most appropriate technology to use and accelerate their implementation.

In practice a strong overlap can exist in early careers, as many RSEs are hired after a physics PhD (although this is very uneven across regions). The careers are, however, frequently differentiated, with different criteria for career advancement. Domain specific software experts, for example, are often evaluated primarily on the production of scientific papers, while their primary activity is related to software development. On the other hand, limited freedom is sometimes provided to RSEs assigned to supporting services rather than research groups. In both cases, the gap should be closed by training in specific software skills for DSSEs and scientific culture for RSEs. Making clear what role software specialists play in our scientific endeavour is essential to create a sense of belonging and maintain engagement.

Hiring a mixture of people, from different backgrounds, and having them work closely as a \emph{team} on common problems, will lead to a very positive \emph{interdisciplinary} approach with the best outcomes for research software and, thus, for our science. In a time when workplace culture, sense of purpose, and recognition play a bigger role than salary in job choices, the community has a unique opportunity to attract new talent to the field and prepare for the future.

\subsection{Sustaining Training Events and Material}

As mentioned several times in this document, and discussed extensively in JENA WP5~\cite{JENA_WG_Reports}, training is the cornerstone for realising the full potential of the APPEC, ECFA and NuPPEC scientific communities. Varied and abundant training material (suitable for self-study) should therefore be produced and curated, and it is therefore essential that training activities are recognised as an essential part of research.

A mixture of in-person and online training schools, self-led training material, and of course best practice guidelines should be produced. There is a strong opportunity for sharing the training effort between our three sub-communities as many of the challenges and tools are the same. Networks such as EVERSE are key in delivering high quality material at a lower human cost by bringing together a large panel of software specialists.

\section{Conclusion and Recommendations}

The research communities of APPEC, ECFA and NuPPEC face significant challenges related to software and computing in the coming years. From the JENA-Spectrum survey 90\% of respondents reported that improving their software performance is an issue, due to data rates, physics goals and resource shortfall; additionally 58\% reported lacking the necessary resources to do so. Here we identify specific positive policy decisions that, backed up by investment, will help.

{\bf 1. Develop strong software policies aligned with Open Science and FAIR Principles and provide resources to support those policies}

Only with appropriate software is the value of open data unlocked, so FAIR4RS principles should be expected and rewarded. Clear policies and support from laboratories and host institutes can help researchers in multiple areas to build up a body of practice from which everyone will benefit. This practice should be based on, and develop, appropriate standards with community and expert input.

We make note in particular of the need to support access to important previous results in our fields, such as HEPdata or the proper management of nuclear physics data, which is a combined software and data preservation and access challenge requiring ongoing investment.

{\bf 2. Create Software Management Plans~\cite{SMPs} and road maps to identify critical software that should be supported by adequate investment in software maintenance, refactoring, and development of modern solutions that help both reduce environmental impact and tackle new scientific challenges}

The maintenance of current software in the face of system upgrades and architectural diversity needs low level, but consistent, support from experts. In parallel, there is also a great need for radically improved software vital to manage the data complexity and rates from future facilities.

Modern hardware demands the use of parallelism for efficiency, where multi-threading is usually also required in order to manage memory consumption. For the most intensive tasks, moving to SIMD and GPU processing can bring huge benefits, but the development costs and required skill levels are significant. As platforms diversify, more effort is then needed to maintain, build and validate systems. To make use of the heterogeneous resources expected at European computing sites, such an investment is vital. Ensuring a vigorous program of R\&D into promising languages (e.g., Julia) and into abstraction layers in our workhorses of Python and C++ will enable informed choices that will ultimately save effort.

Investment in efficient, flexible code not only can give better science outcomes, but has a direct impact on reducing the environmental costs of computing: fewer machines are needed if the same science can be achieved using more algorithmically efficient software running on more power efficient platforms.

{\bf 3. Strategically invest in software that serves multiple experiments and disciplines, and which optimises data and workflow management}

Insofar as software can meet common goals for multiple users, well engineered systems deserve strategic investment, particularly given the software engineering demands above. Common, well-written software is particularly valuable for smaller experiments who may struggle to maintain even their basic software stack.
Of particular note, unlocking the potential of distributed computing, especially for smaller endeavours, needs better, easier to use tools for data and workflow management.
Collaborations across disciplines need specific targeted support so that all members can effectively contribute and benefit. We note existing initiatives such as EOSC (e.g., the EVERSE project), the Science Clusters~\cite{ScienceClusters}, such as ESCAPE, and community efforts such as the HEP Software Foundation~\cite{HSF}.

{\bf 4. Support the optimal use and allocation of computational resources}

Notwithstanding software improvements, experiments will require improved computational resources to support higher data rates, the growing need for real-time data processing, and complex data. This may involve upgrading existing computing infrastructure and exploring, e.g., cloud-based solutions, and must be done as part of a dialogue between the software experts and facilities. Integration of these multi-faceted solutions requires software and computing expert effort to support the optimal, sustainable use of computing resources. This effort must be seen as an integral part of the design, construction and exploitation of new experiments.

{\bf 5. Invest in training and reward trainers, adapting to new technologies and techniques as they arise}

The range of skills required to develop effective, scalable software solutions is very broad, with machine learning, data science techniques, high-performance computing and DevOps all playing a role. With new technologies continuing to emerge both now and for the foreseeable future, this requires quality training material, written by experts and presented by expert instructors. Such significant effort needs to be properly recognised as being highly valuable and rewarded appropriately so that people will be motivated to develop, support and teach these skills.

{\bf 6. Reward software and computing work and provide suitable career paths}

To achieve our science goals in a way that meets the requirements of Open Science and satisfies FAIR principles requires investment in people who can write, maintain and improve the software that is required by our collaborations. Such work needs to be seen as an integral part of our experiments and a first class citizen of research.  Investing in high quality software that can be maintained and reused is the only realistic path to achieving the goals of Open Science while simultaneously supporting the operational needs of the experiments.

Crucially, this investment has to cover viable long-term career paths for such staff, to allow them to support their software for as long as needed, which in the case of Open Science is as long as possible.
New roles are emerging that can be considered as viable for software experts -- in particular the research software engineer position. Policy makers and funders should continue to be creative in finding the means to support the software experts on whom our science so crucially relies.

\section*{References}
\printbibliography[heading=none]

\end{document}